# Introducing a New Mechanism for Construction of an Efficient Search Model

**Debajyoti Mukhopadhyay[1,3], Sukanta Sinha[2,3]**
[1] Balaji Institute of Telecom & Management, Pune411033 , India
[2] Tata Consultancy Services,Victoria Park, Kolkata 700091, India
[3] WIDiCoReL, Green Tower C- 9/1, Golf Green, Kolkata 700095, India
*{ debajyoti.mukhopadhyay, sukantasinha2003}@gmail.com*

*ABSTRACT*
*Search engine has become an inevitable tool for retrieving information from the WWW. Web researchers introduce lots of algorithms to modify search engine based on different features. Sometimes those algorithms are domain related, sometimes they are Web page ranking related, and sometimes they are efficiency related and so on. We are introducing such a type of algorithm which is multiple domains as well as efficiency related. In this paper, we are providing multilevel indexing on top of Index Based Acyclic Graph (IBAG) which support multiple Ontologies as well as reduce search time. IBAG contains only domains related pages and are constructed from Relevant Page Graph (RPaG). We have also provided a comparative study of time complexity for the various models.*

## 1. INTRODUCTION
Search engine retrieves document from World Wide Web (WWW) [1][2]. Typically a search engine does not support domain related searching. We introduce single domain specific search engine and multiple domains specific search engine for getting more prominent search results.

RPaG model consists of multiple domain specific Web-pages [3][4][5][6]. This model takes huge time to retrieve the data when a search has been made based on the specified model specially for handling large data storage. In this background, we incorporate a new IBAG model which provides faster access of Web pages to the users [7][8][9][10].

IBAG is constructed based on Web-page mean relevance range which is determined by maximum mean relevance span value, minimum mean relevance span value and number of mean relevance span level [7]. Now consider a scenario where, for a single relevance range huge number of Web-pages exists, i.e., huge number of Web-pages exists whose mean relevance value is approximately the same and belongs to the same level. In this case, searching a list of Web-pages from IBAG will take huge time. To overcome this situation we introduce a new model called Multilevel Index Based Acyclic Graph (M-IBAG) which supports multilevel indexing.

This paper involves the basic idea of searching Web-pages from IBAG and RPaG models. Also describes a design and development methodology for construction of M-IBAG from IBAG. Earlier we constructed IBAG from RPaG. RPaG is generated based on multiple Ontologies, for this IBAG also support multiple Ontologies [11][12][13][14]. Finally, we provide a comparative study for searching a Web-page from different models.

## 2. EXISTING MODEL
In this section, we describe two existing models - RPaG model and IBAG model. RPaG is generated while original crawling happens and then using a typical mechanism IBAG generated from that RPaG.

**Definition 1.** Weight Table - This table contains two columns; first column denotes Ontology terms and second column denotes weight value of that Ontology term. Weight value must be within '0' and '1'..

**Definition 2.** Syntable - This table contains two columns; first column denotes Ontology terms and second column denotes synonym of that ontology term. For a particular ontology term, if more than one synonyms exists then it should be kept using comma (,) separator.

### 2.1. RELEVANCE PAGE GRAPH MODEL
In this section, RPaG is described along with the concept of its generation procedure. Every Crawler [3][4][5][6][15][16][17] needs some seed URLs to retrieve Web-pages from World Wide Web (WWW). All Ontologies, Weight Tables and Syntables [18] are needed for retrieval of relevant Web-pages. RPaG is generated only considering relevant Web-pages. Each node in RPaG holds Web-page information. In RPaG, each node contains P_ID, URL, PP_ID1, PP_ID2, PP_ID3, PP_ID4, ONT_1_REL_VAL, ONT_2_REL_VAL, ONT_3_REL_VAL, ONT_1_F, ONT_2_F and ONT_3_F field information. Suppose we select a Web-page P. In

RPaG, all field values of Page P are stored. P_ID is Page Identifier of Page P, which is a unique number for each page. PP_IDs are Parent Page Identifier of Page P. we are taken four PP_ID, just for improving accuracy. 'Ontology 1' relevance value (ONT_1_REL_VAL) of Page P is generated according to the 'Ontology 1'. Similarly, 'Ontology 2' relevance value (ONT_2_REL_VAL) of Page P is generated according to the 'Ontology 2'. Again, 'Ontology 3' relevance value (ONT_3_REL_VAL) of Page P is generated according to the 'Ontology 3'. If Page P supports 'Ontology 1'; i.e., relevance value overcomes relevance limit; then 'Ontology 1' flag (ONT_1_F) must be 'Y'. If Page P supports 'Ontology 2'; i.e., relevance value overcomes relevance limit; then 'Ontology 1' flag (ONT_2_F) must be 'Y'. If Page P supports 'Ontology 3'; i.e., relevance value overcomes relevance limit; then 'Ontology 1' flag (ONT_3_F) must be 'Y'. A sample RPaG is shown in Fig. 2. Each node in this figure of RPaG contains four fields; i.e., Web-page URL, ONT_1_REL_VAL, ONT_2_REL_VAL and ONT_3_REL_VAL.

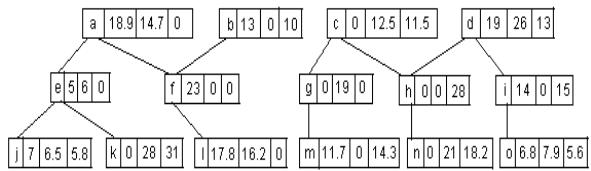

**Figure 1.** A Simple Example of Relevance Page Graph

## 2.2. INDEXED BASED ACYCLIC GRAPH MODEL

IBAG is multiple Ontology supported model and constructed from RPaG [3][7]. RPaG pages are related in some Ontologies and the IBAG generated from this specific RPaG is also related to the same Ontologies. Each node of IBAG contains Page Identifier (P_ID), Unified Resource Locator (URL), Parent Page Identifier (PP_ID), Mean Relevance value (MEAN_REL_VAL), Ontology 1 link (ONT_1_L), Ontology 2 link (ONT_2_L) and Ontology 3 link (ONT_3_L) fields. P_ID is selected from RPaG Page Repository. Each URL has a unique P_ID and the same P_ID of the corresponding URL is mentioned into IBAG page repository. Consider, one page supports 'Ontology 1' and 'Ontology 2'; then we calculate MEAN_REL_VAL as (ONT_1_REL_VAL + ONT_2_REL_VAL)/2. If one page supports 'Ontology 1', 'Ontology 2' and 'Ontology 3'; then we calculate MEAN_REL_VAL as (ONT_1_REL_VAL + ONT_2_REL_VAL + ONT_3_REL_VAL) / 3. 'Ontology 1 link' (ONT_1_L), 'Ontology 2 link' (ONT_2_L) and 'Ontology 3 link' (ONT_3_L) points to the next 'Ontology 1', 'Ontology 2' and 'Ontology 3' supported pages respectively. In Fig. 2, we have shown only four fields; i.e., page URL, ONT_1_L, ONT_2_L and ONT_3_L.

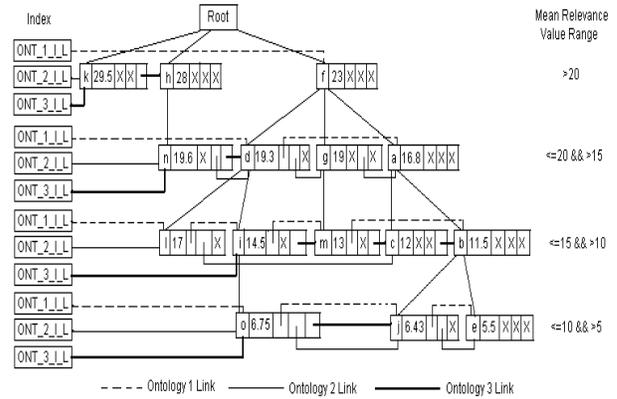

**Figure 2.** IBAG Model for the Above Given RPaG

Construction of IBAG demands "Maximum Mean Relevance Span Value" (α), "Minimum Mean Relevance Span Value" (β) and "Number of Mean Relevance Span level" (m) as input to define mean relevance value range of each level. It is calculated using the formula given below:

$$\text{Mean Gap Factor } (\rho) = (\alpha - \beta) / m$$

Now we define ranges such as β to β+ ρ, β+ ρ to β+ 2ρ, β+ 2ρ to β+ 3ρ and so on. In each level, all the Web-pages' "Mean Relevance Value" are kept in a sorted order and all the indexes which track that domain related pages are also stored.

## 3. STUDY OF EXISTING MODELS

In this section we have describe Web-page traversing technique from RPaG and IBAG model. Also describe why we introduce multilevel indexing mechanism.

### 3.1. RPaG TRAVERSING TECHNIQUE

In existing RPaG model, Web-page searching technique follows linear search mechanism. Every time searching starts from the starting Web-page. Suppose, we would like to search one Web-page 'm' from the RPaG model. Web-page 'm' supports 'Ontology 1' and 'Ontology 3' because 'Ontology 2' relevance value is '0'. Now, the Web-page would be definitely read by traversing Web-pages one by one irrespective of their supported Ontologies and traversing starts from starting Web-page (here Web-page 'a'). Reading mechanism of RPaG is shown in Figure 3.

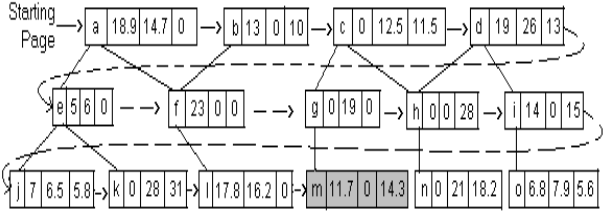

**Figure 3.** RPaG Reading Mechanism

### 3.2. IBAG TRAVERSING TECHNIQUE

Existing IBAG model supports three Ontologies; hence all the level starts with three Ontology Indexes. Each level 'Ontology 1 Index' linked with first 'Ontology 1' supported page of that level. Each level 'Ontology 2 Index' linked with first 'Ontology 2' supported page of that level. Each level 'Ontology 3 Index' linked with first 'Ontology 3' supported page of that level. All pages in IBAG contain three link fields. First one for 'Ontology1',

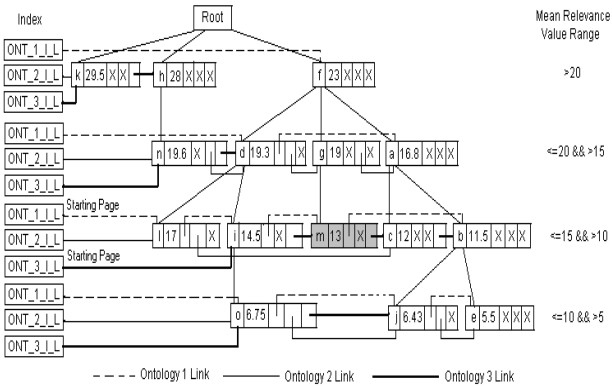

**Figure 4.** IBAG Reading Mechanism

Second one for 'Ontology 2' and Third one for 'Ontology 3'. Now, if any page supports all three Ontologies then we traverse next page through that page. Again if any page supports the 'Ontology 1' and 'Ontology 3'; then we traverse next 'Ontology 1' supported page and 'Ontology 3' supported page through that page. Say, we would like to search one Web-page 'm' from the IBAG model. Web-page 'm' supports 'Ontology 1' and 'Ontology 3' and belongs to level 3. Now, the Web-page would be definitely read at level 3 starting with 'Ontology 1 Index' and 'Ontology 3 Index'. Reading mechanism of IBAG is shown in Figure 4.

### 3.3. REASON OF INTRODUCING MULTILEVEL INDEXING CONCEPT

Consider an IBAG model which contains 'n' number of Web-pages and 'm' number of levels. Now for an ideal scenario, IBAG model contains (n/m) number of Web-pages in each level. An ideally distributed Web-pages IBAG is shown in Figure 5.

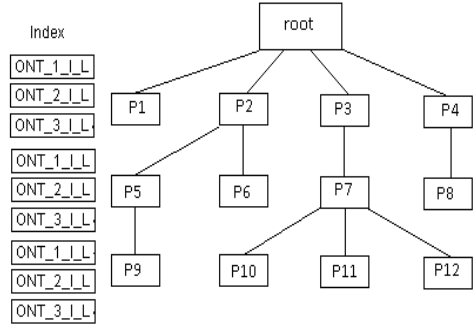

**Figure 5.** IBAG (Ideal Case)

Again we consider the scenario like $(n/m) \approx n$ i.e. all Web-pages belongs to a single level, shown in Figure 6. While retrieve a Web-page from the Figure 6, we are not getting any benefit in the search time prospective which is described in section 5.3. To reduce search time for this type of scenario we introduced a new model multilevel IBAG which is briefly explained in section 4.

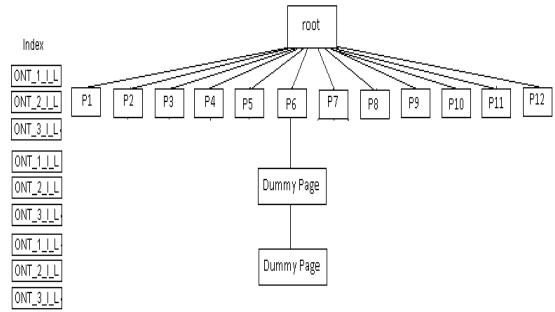

**Figure 6.** IBAG (While $(n/m) \approx n$)

### 4. PROPOSED APPROACH

In our approach, we have generated a new model M-IBAG, which produced better time complexity under any circumstance. Section 4.1 explains M-IBAG model working principle and section 4.2 depicts construction mechanism of M-IBAG model for a given IBAG model.

#### 4.1. M-IBAG Model

In IBAG model, number of Web-page exists in a particular mean relevance level doesn't matter. Where as we were more focusing on the Web-pages belongs to which mean relevance range. For this reason, if all Web-pages mean relevance value belongs to a single mean relevance range then all the Web-pages exists in a single level (refer Fig. 6). Now while we search Web-pages from this type of IBAG model for a user given search string, this prototype should not produce better search time complexity with respect to RPAG model search time complexity. To overcome this type of situation we introduce a new model which is an extended version of IBAG model named as M-IBAG model.

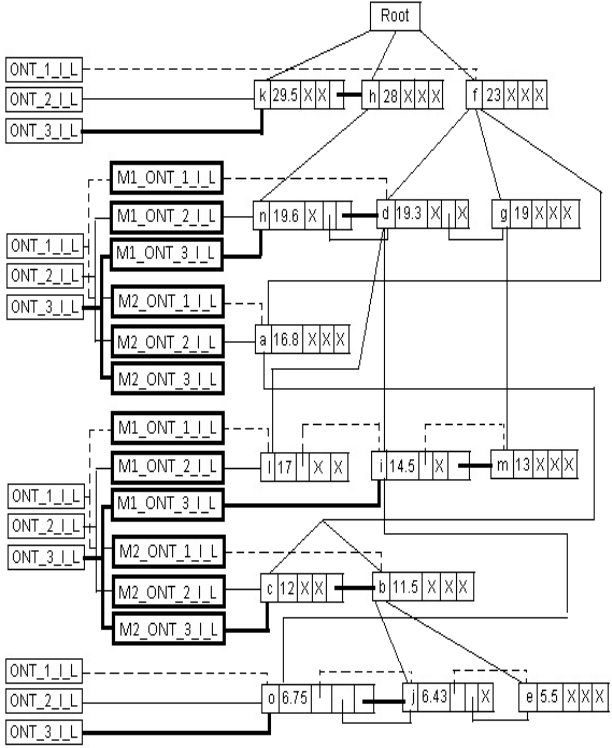

**Figure 7.** M-IBAG Model

The construction mechanism of M-IBAG model in each level should not exceed 'Floor of (n/m)' number of Web-pages, where n and m denotes total number of Web-pages in IBAG and number of mean relevance range span respectively. In IBAG model, if any mean relevance level contains more than 'Floor of (n/m)' number of Web-pages then we apply multilevel concept and ensures that each level maximum contain 'Floor of (n/m)' number of Web-pages otherwise we leave the level as it is. In Figure 7 a sample M-IBAG is shown. IBAG Web-pages are related to some Ontology, and the M-IBAG generated from this specified IBAG is also related to the same Ontologies.

### 4.2. M-IBAG CONSTRUCTION

To construct multilevel indexed IBAG, our proposed algorithm expecting IBAG as an input. Here we describe our approach in a brief using pseudo code. We assume that in IBAG contains 'n' number of Web-pages and those 'n' numbers of Web-pages were distributed in 'm' number of Mean Relevance Range Spans.

**ConstructionOfMultilevelIndexedIBAG**(IBAG)
*begin*
 *do for ever*
   *wait (ConstructionOfMultilevelIndexedIBAG);*
*while (empty (IBAG))*
*begin*
  *find Multilevel Web-page Limit (nMultiLvlLmt) :=*
  ⌊*(n/m)*⌋ *i.e. Floor of (n/m);*
  *begin*
    *for each Mean Relevance Range level*
     *find count Number of Web-pages exists (μ) in*
     *the current Mean Relevance Range level;*
     *if (μ > nMultiLvlLmt )*
      *call MultiLevelIndexing(IBAG, Current Mean*
      *Relevance Range Level, μ, nMultiLvlLmt);*
     *else*
      *do nothing;*
    *End;*
   *end loop;*
  *end;*
*end;*
 *signal (hungry);*
*end.*

**MultiLevelIndexing**(IBAG, Current Mean Relevance Range Level, Number of Web-pages exists, Number of Web-page Limit for Multilevel)

*begin*
 *do for ever*
   *wait (MultiLevelIndexing);*
*while (empty (IBAG))*
*begin*
  *Find Number of Indexing Required (η):=*
  *(μ / nMultiLvlLmt);*
  *Find Ceiling of "Number of Indexing Required (η)"*
  *i.e.* ⌈η⌉ *;*
  *begin*
   *do for ever*
    *start Index Level Count from 1;*
    *begin*
     *do for ever*
      *start Number of Web-page Traversed from 1;*
      *add the traversed Web-page with current*
      *iIndex level and update Index links;*
      *increase Number of Web-page Traversed by 1;*
     *while (Number of Web-page Traversed <=*
        *nMultiLvlLmt);*
    *end;*
    *increase Index Level Count by 1;*
    *while (Index Level Count <=* ⌈η⌉ *);*
  *end;*
*end;*
 *signal (hungry);*

*end.*

## 5. TIME COMPLEXITY ANALYSIS FOR DIFFERENT MODELS

In this section we are giving some comparative study of time complexity for RPaG model, both IBAG model i.e. ideal case and while all the Web-pages belongs to same level and M-IBAG model. We assume that each model contains 'n' number of Web-pages or nodes and also assume that to traverse a single Web-page or Node or an Index takes 1 millisecond time. Below given time complexity calculations for retrieving a single Web-page from different models.

### 5.1. RPaG MODEL

In RPaG model while search a Web-page, traversing always starts from a single position and then follow linear searching mechanism. An arbitrary example of an RPaG model is shown in Figure 1.

**Best Case Time Complexity.** Best case scenario of RPaG model is first Web-page of the traversing link should be our resultant page. Number of traversed Web-page should be 0 for finding the resultant Web-page. Hence the best case complexity becomes a constant (say c) for search a Web-page from RPaG model. The straight forward solution takes constant time, which we write as $O(1)$.

**Worse Case Time Complexity.** Worse case scenario of RPaG model is last Web-page of the traversing link should be our resultant page. To find the resultant Web-page we need to traverse (n-1) number of Web-pages. And (n-1) always less than n for all n>0. Hence the worse case complexity become $O(n)$ for search a Web-page from RPaG Model.

**Average Case Time Complexity.** We have 'n' number of Web-pages. To search all Web-pages from RPaG model we need $[0+1+2+3+ \ldots + (n-1)]$ msec. Now for finding a single Web-page from PRaG model requires $1/n * [0+1+2+3+ \ldots + (n-1)]$ msec. Hence the average case complexity becomes

$$1/n * [0+1+2+3+ \ldots + (n-1)]$$
$$= 1/n * [(n-1)(n-1+1)/2]$$
$$= (n-1)/2 < n \ \forall \ n>0$$
$$\approx O(n)$$

### 5.2. IBAG MODEL: IDEAL CASE

In IBAG model all the Web-pages distributed in different mean relevance levels. Now, we assume that 'n' numbers of Web-pages are distributed in 'm' number of Mean Relevance Level and all the Web-pages support all three Ontologies. For an Ideal IBAG model each level contains (n/m) number of Web-pages i.e. Web-pages are equally distributed in each level. Time taken for retrieve a single Web-page from ideal IBAG model for the various scenarios will describe below. Figure 5 have shown an ideal IBAG model.

**Best Case Time Complexity.** In best case of an Ideal IBAG model found the resultant Web-page just traversing only one index node. Hence the best case time complexity becomes a constant i.e. $O(1)$ for retrieve a single Web-page from ideal IBAG model.

**Worse Case Time Complexity.** For an ideal IBAG model each level contains n/m number of Web-pages. Now to get the last Web-page from any one of the level in ideal IBAG model, we need to traverse one index node and (n/m - 1) number of Web-pages. Hence the worse case time complexity become $[1 + (n/m - 1)] = n/m \approx O(n/m)$.

**Average Case Time Complexity.** To retrieve all the Web-pages in a particular level from ideal IBAG model we need to traverse $[(1+0) + (1+1) + (1+2) + \ldots + (1+ (n/m – 1))] = [1+2+3+ \ldots + n/m]$ number of Web-pages. Now for all Web-pages from ideal IBAG model we need to traverse $[1 + 2 + 3 + \ldots + n/m] + [1 + 2 + 3 + \ldots + n/m] + [1 + 2 + 3 + \ldots + n/m] + \ldots$ m times. Hence to find a single Web-page from ideal IBAG model in average scenario has given below:

$$(1/n) \sum_{Level=1}^{m} [1 + 2 + 3 + \ldots + (n/m)]$$
$$= (1/n) \sum_{Level=1}^{m} [((n/m)*(n/m + 1))/2]$$
$$= \sum_{Level=1}^{m} [(n/m + 1)/(2*m)]$$
$$= m*[(n/m + 1)/(2*m)]$$
$$= (n/m + 1)/2 < (n/m) \ \forall \ n>0, m>0 \text{ and } n>m$$
$$\approx O(n/m).$$

### 5.3. IBAG MODEL: WHILE ALL THE WEB-PAGES BELONGS TO SAME LEVEL

In an IBAG model, all the Web-pages belong to same level means all 'n' number of Web-pages exists in same level. For this type of situation, IBAG model can't produce better time complexity with respect to RPaG model. A sample scenario is shown in Figure 6. Time taken for retrieving a single Web-page from this type of model is given below.

**Best Case Time Complexity.** In best case scenario, we need to traverse index node and then we will get the Web-page. Hence the time complexity becomes a constant i.e. $O(1)$ for retrieve a single Web-page from this IBAG model.

**Worse Case Time Complexity.** For this IBAG model, in worse case, we have to find the last Web-page of that particular level where all the Web-pages exist. Now to find the last Web-page we need to traverse one Index node and (n-1) number of Web-pages. Hence the worse case time complexity become $[1 + (n-1)] = n \approx O(n)$.

**Average Case Time Complexity.** In average case scenario first we were calculating time for retrieve all Web-pages from this IBAG model. Then we were taking average to find a single Web-page. While calculating time we need to consider 1 msec taken for traversing index node and then we count how many Web-pages need to be traverse to get all Web-pages. Average case time taken for finding one single page from this IBAG model given below:

$$(1/n) [(1+0)+(1+1)+(1+2)+ \ldots +(1+(n-1))]$$
$$= (1/n) [1+2+3+\ldots+n]$$
$$= (1/n) [n*(n+1)/2]$$
$$= (n+1)/2 < n \ \forall n > 0$$
$$\approx O(n)$$

### 5.4. M-IBAG MODEL

In Figure 6 have shown a sample M-IBAG model. Now based on our M-IBAG construction mechanism, each level of M-IBAG should contain maximum n/m number of Web-pages, where n and m denotes total number of Web-pages and number of mean relevance level respectively. To find a Web-page from M-IBAG first we have to traverse index then if multilevel index available then we traverse the multilevel index and then the corresponding Web-page links otherwise we directly traverse Web-page links.

**Best Case Time Complexity.** In best case scenario to retrieve a Web-page from M-IBAG model we need to traverse only one index and get the Web-page. We assume, that particular level not eligible for multilevel i.e. already contain less than or equal to n/m number of Web-pages. Hence the best case complexity for search a Web-page from M-IBAG model becomes a constant and denoted as O(1).

**Worse Case Time Complexity.** Each level of M-IBAG model maximum contains n/m number of Web-pages; hence to fetch one Web-page we need at most (n/m-1) number of Web-pages. For the worse case scenario, to retrieve a Web-page from M-IBAG model we have to traverse one Index and then one multilevel index and then (n/m-1) number of Web-pages. Hence the worse case complexity for search a Web-page from M-IBAG model becomes

$$[1 + 1 + (n/m - 1)] = (1+n/m)$$
$$\leq 2*(n/m) \ \forall n > 0, m > 0 \text{ and } n > m$$
$$\approx O(n/m)$$

**Average Case Time Complexity.** We have 'm' number of mean relevance level. Out of 'm' number of mean relevance level at most we have to allow (m-1) number of mean relevance level for multilevel indexing because our total number of Web-pages is fixed. For each level search all Web-pages we have to traverse maximum [1+2+3+ … + (n/m)] number of Web-pages if multilevel indexing have present. Now for the m number of level we need to traverse $\sum_{Level=1}^{m}[1+2+3+\ldots+(n/m)]$ number of Web-pages and for multilevel index we have to traverse at most (m-1) number of multilevel index. Hence to search a single Web-page from M-IBAG model for the average case scenario takes $(1/n) [(m-1) + \sum_{Level=1}^{m}[1+2+3+\ldots+(n/m)]]$ msec. Average case time complexity derivation given below:

$$(1/n) [(m-1) + \sum_{Level=1}^{m}[1+2+3+\ldots+(n/m)]]$$
$$= (m-1)/n + (1/n) \sum_{Level=1}^{m} [((n/m)*(n/m+1))/2]$$
$$= (m-1)/n + \sum_{Level=1}^{m} [(n/m+1)/2*m]$$
$$= (m-1)/n + [m*(n/m+1)/2*m]$$
$$= (m-1)/n + (n/m+1)/2$$

Now, (m-1)/n always less then 1 because under any circumstance m < n. Hence, if we neglect (m-1)/n part then the average case complexity of M-IBAG model under any circumstance becomes

$$(n/m+1)/2 \leq (n/m) \ \forall n > 0, m > 0 \text{ and } n > m$$
$$\approx O(n/m)$$

### 5.5. COMPARATIVE STUDY OF TIME COMPLEXITY FOR THE ABOVE GIVEN MODELS

In this section we have described the comparative study of time complexity for RPaG model, both IBAG model i.e. ideal case and while all the Web-pages belongs to same level and M-IBAG model. And we found that under any circumstance M-IBAG model gives better time complexity while retrieving a Web-page.

**Table 1.** Comparative Study of Time Complexity

| Case | RPaG Model | IBAG Model (Ideal Case) | IBAG Model (Where (n/m) ≈ n) | M-IBAG Model |
|---|---|---|---|---|
| Best | O(1) | O(1) | O(1) | O(1) |
| Worse | O(n) | O(n/m) | O(n) | O(n/m) |
| Average | O(n) | O(n/m) | O(n) | O(n/m) |

## 6. CONCLUSIONS

In this paper, we proposed a prototype of a domain specific Web search engine which supports multiple Ontologies. This prototype has an enhanced version of IBAG model. While searching Web-pages based on the user given search string our prototype retrieves Web-pages from M-IBAG model. This prototype has not only produced faster result but also it is highly scalable. We increase domain only by introducing new domain Ontology. Overall, the proposed algorithm has shown the construction of the M-IBAG model. Also given some comparative study of time complexity for RPaG model, both IBAG model (i.e., ideal case and while all the Web-pages belongs to same level), and M-IBAG model.